\documentclass[aps, reprint,showpacs,showkeys,prl,superscriptaddress]{revtex4-2}
\usepackage{dcolumn}
\usepackage{comment} 
\usepackage[utf8]{inputenc}
\usepackage{amsmath}
\usepackage{amsfonts}
\usepackage{amssymb}
\usepackage{braket}
\usepackage{graphicx}
\usepackage{hyperref}
\usepackage{xcolor}
\usepackage{braket}
\usepackage{siunitx}

\DeclareMathAlphabet{\mathdutchcal}{U}{dutchcal}{m}{n}
\SetMathAlphabet{\mathdutchcal}{bold}{U}{dutchcal}{b}{n}

\newcommand*{\QOQI}{Quantum Optics and Quantum Information Group, \\Friedrich-Alexander-Universität Erlangen-Nürnberg, Staudtstr. 1, 91058 Erlangen, Germany}

\begin{document}

\title{Multi-Path and Multi-Particle Tests of Complex vs. Hyper-Complex Quantum Theory}

\author{Ece \.Ipek Saruhan}
\affiliation{Institute for Quantum Optics and Quantum Information - IQOQI Vienna, Austrian Academy of Sciences, Boltzmanngasse 3, A-1090 Vienna, Austria}
\affiliation{\QOQI}
  
\author{Joachim von Zanthier}
\affiliation{\QOQI}

\author{Marc-Oliver Pleinert}
\affiliation{\QOQI}

\begin{abstract}
The axioms of quantum mechanics provide limited information regarding the structure of the Hilbert space, such as the underlying number system. The latter is generally regarded as complex, but generalizations of complex numbers, so-called hyper-complex numbers, cannot be ruled out in theory. Therefore, specialized experiments to test for hyper-complex quantum mechanics are needed. To date, experimental tests are limited to single-particle interference exploiting a closed phase relation in a three-path interferometer called the Peres test. The latter distinguishes complex quantum mechanics from quaternionic quantum mechanics. Here, we propose a general matrix formalism putting the Peres test on a solid mathematical ground. On this basis, we introduce multi-path and multi-particle interference tests, which provide a direct probe for any dimension of the number system of quantum mechanics.

\end{abstract}

\maketitle

Modern quantum mechanics -- formulated 100 years ago~\cite{Heisenberg:1925,Schrodinger:1926} -- has sparked unique curiosity in science due to its counter-intuitive predictions: 
A single particle can be in a superposition state and thus interfere with itself~\cite{Feynman:2010c,Eichmann:1993,Arndt:1999}; moreover, two particles can be entangled giving rise to nonlocality in terms of Bell inequality violations~\cite{Bell:1964,Freedman:1972,Aspect:1982,Weihs:1998}. 
These \textit{``mysteries''}~\cite{Feynman:2010c} have led to questioning quantum mechanics from the start, even by its founders.
Schrödinger, who introduced the complex number $i$ in his famous equation for the dynamics of the wave function $\psi$~\cite{Schrodinger:1926}, later criticized its use, \textit{``What is unpleasant here, and indeed directly to be objected to, is the use of complex numbers. $\psi$ is surely fundamentally a real function''}~\cite{Einstein:2011}.
Nowadays, the foundational pillars of quantum mechanics, i.e., its axioms, are typically challenged in specialized tests~\cite{Sorkin:1994,Sinha:2010,Kauten:2017,Cotter:2017,Vogl:2021,Pleinert:2020,Pleinert:2021,Renou:2021,Li:2022,Chen:2022,Peres:1979,Kaiser:1984,Procopio:2017,Gstir:2021,Sadana:2022}, also to rule out alternatives to quantum theory~\cite{Stueckelberg:1960,McKague:2009,Aleksandrova:2013,Birkhoff:1936,Jordan:1933,Jordan:1934,Finkelstein:1962,Finkelstein:1963,Horwitz:1984,Horwitz:1996,Adler:1995,Gunaydin:1973,Gunaydin:1974,Leo:1996,Mironov:2015,Dakic:2014}.

One of the very axioms of quantum mechanics is Born's rule~\cite{Born:1926}, which establishes a connection between the abstract mathematical formalism and actual experiments. It states that it is the absolute square of the complex\mbox{(-valued)} quantum-mechanical wave function $\psi$ that is related to the real\mbox{(-valued)} world, partly resolving Schrödinger's ``complexity issue''.
Initiated by Sorkin~\cite{Sorkin:1994}, the rule has been subject to several single-particle tests using various platforms over the last 15 years~\cite{Sinha:2010,Kauten:2017,Cotter:2017,Vogl:2021}.
Recently, these tests were extended to multi-particle verifications of Born's rule which exhibit a higher sensitivity to deviations~\cite{Pleinert:2020} and a first two-particle test of Born's rule was conducted~\cite{Pleinert:2021}.
So far, no deviations from the rule have been found.

A further building block of quantum theory is the continued use of complex numbers in quantum mechanics (CQM). 
However, other quantum-mechanical formulations -- based on different number systems like real numbers or even hyper-complex numbers -- are technically possible.
Real-valued quantum mechanics (RQM), for instance, has been explored~\cite{Stueckelberg:1960,McKague:2009,Aleksandrova:2013}. 
Only recently, Renou et al. proposed a Bell-like experiment to show the shortcoming of a solely real-valued theory~\cite{Renou:2021}. The latter has been experimentally verified a year later with photons~\cite{Li:2022} and superconducting qubits~\cite{Chen:2022}.
Complex numbers are therefore necessary for quantum theory, but the question remains as to whether they are sufficient.
Attention has therefore turned to higher-dimensional formulations.

The use of quaternions (4-dimensional generalizations of complex numbers) and even octonions (8-dimensional generalizations) was already discussed in the early days of quantum mechanics by Birkhoff and von Neumann~\cite{Birkhoff:1936}, and by Jordan, von Neumann, and Wigner~\cite{Jordan:1933,Jordan:1934}, respectively. 
Later, the foundations of a quaternionic quantum mechanics (QQM) were further explored~\cite{Finkelstein:1962,Finkelstein:1963,Horwitz:1984,Horwitz:1996}, while a comprehensive study on QQM was published by Adler in 1995~\cite{Adler:1995}. 
Octonionic~\cite{Gunaydin:1973,Gunaydin:1974,Leo:1996} and even sedenionic~\cite{Mironov:2015} constructions have also been investigated, e.g., in attempts to explain quark confinement~\cite{Gunaydin:1974} or the generalization of quantum mechanics and field theory equations~\cite{Mironov:2015}.
Note that extending quantum mechanics to higher-dimensional number systems might lead to mathematical inconsistencies such as non-associativity. 
Yet, the latter, for instance, has been addressed by using complex geometry~\cite{Leo:1996}. 

A first test to differentiate between CQM and QQM experimentally was proposed by Peres in 1979 utilizing single-particle interference in a three-path setup~\cite{Peres:1979}.
Early experimental realizations, conducted with neutrons~\cite{Kaiser:1984} and photons~\cite{Procopio:2017}, utilized a trimmed version of the test. Only recently, measurements of the original Peres test in the optical~\cite{Gstir:2021} and microwave regime~\cite{Sadana:2022} were realized, yet both unable to rule out QQM. 
Although motivated by Adler already in 1995,
"\textit{to provide tests for quaternionic quantum mechanics [...] A potentially fruitful avenue, which has not yet been explored, is that of multiparticle effects.}" \cite{Adler:1995}, there has been so far no work on the extension of the Peres test to multi-path and multi-particle interference.

In this Letter, we address this issue. 
We first briefly recapitulate quaternions as an example for a number system of a hyper-complex quantum-mechanical theory and the Peres test.
We then introduce a matrix method which allows to recover the single-particle Peres test in a general frame, also revealing a geometric interpretation of the test. 
Based on this, we generalise the single-particle Peres test to an arbitrary number of paths and particles revealing a direct link to the dimension of the number system.

\paragraph*{Quaternions.}

The most prominent example of the construction of a hyper-complex theory beyond standard quantum mechanics, is the four-dimensional formulation of quantum mechanics based on quaternionic wave functions~\cite{Adler:1995}. 
The Hilbert space $\mathcal{H}$ is in that case quaternionic $\mathbb{H}$, and an element of that space, a quaternion $q\in\mathbb{H}$, is an extension of a complex number $z=a+b\mathbf{i}\in\mathbb{C}$ to $ q= a + b \mathbf {i} + c \mathbf{j}+ d\mathbf {k}$, where $a,b,c,d$ are real numbers and $\mathbf{i},\mathbf{j},\mathbf{k}$ form the imaginary unit basis with multiplication rules $\mathbf{i}^2 = \mathbf{j}^2 = \mathbf{k}^2 = \mathbf{i}\mathbf{j}\mathbf{k} = -1$.
In general, the multiplication of quaternions is non-commutative, e.g., $\mathbf {i}\mathbf {j} = - \mathbf {j}\mathbf {i} $. 
A quaternion $q$ can also be represented as $q=(v,\mathbf{v})$, i.e., a composition of a scalar part $v=a$ and a pure imaginary vector part $\mathbf{v} = b \mathbf {i} + c \mathbf{j}+ d\mathbf {k}$, analog to the complex decomposition $z=(a,b)$. 
In the same analogy, a quaternion can be expressed in the exponential form $q=|q| e^{\hat{n}\theta}=|q| [\cos (\theta) +\hat{n}\sin (\theta)]$ with  $|q|=\sqrt{a^2+b^2+c^2+d^2}$ being the norm of the quaternion, $\hat{n} = \mathbf{v}/|\mathbf{v}|$ being the unit vector of the pure imaginary part, and 
$|\mathbf{v}|= \sqrt{b^2 + c^2 + d^2}$ being the norm of the vector $\mathbf{v}$ with $\tan(\theta)= |\mathbf{v}|/|a|$.
%

\paragraph*{Peres test.}

In 1979, Asher Peres introduced a method to differentiate between CQM and QQM using a three-path interferometer for single particles~\cite{Peres:1979}. 
An iconic example for such interference is Young's well-known double-slit experiment. Here, each of the two paths is associated to a wave function $\psi_i$ ($i= 1,2$) and the total probability distribution is given according to Born's rule by the absolute square of the superposition of the individual wave functions, i.e., $P_{12}=|\psi_1+\psi_2|^2$.
This formulation leads to interference fringes and 
the (pure) interference term can be extracted by subtracting from the double-slit signal the related single-slit signals $P_i=|\psi_i|^2$, i.e.,
\begin{equation}\label{eq:I_ij}
    \mathcal{I}_{ij} = \frac{P_{ij}- P_i- P_j}{2\sqrt{P_iP_j}}.
\end{equation}
$\mathcal{I}_{ij}$ is called the normalized second-order interference in the terminology of Sorkin's interference hierarchy~\cite{Sorkin:1994,Pleinert:2020}.
In CQM with complex-valued wave functions $\psi_i\propto e^{i\phi_i}$, the interference term corresponds to the cosine of the related phase difference, i.e., $\mathcal{I}^{\,\mathbb{C}}_{ij} = \cos(\phi_{ij}) = \cos(\phi_j-\phi_i)$. 
In QQM with quaternionic-valued wave functions $\psi_i\propto e^{\hat{n}_i\theta_i}$, the interference term becomes
$\mathcal{I}^{\,\mathbb{H}}_{ij} = \cos (\theta_i)  \cos (\theta_j) + \hat{n}_i \cdot \hat{n}_j \sin (\theta_i) \sin (\theta_j)$. 
Since $\mathcal{I}_{ij}^{\,\mathbb{C}/\mathbb{H}} \in [-1,1]$, 
no direct test with two paths can be constructed to differentiate CQM from QQM.

\begin{figure*}
    \centering
    \includegraphics[width=\textwidth]{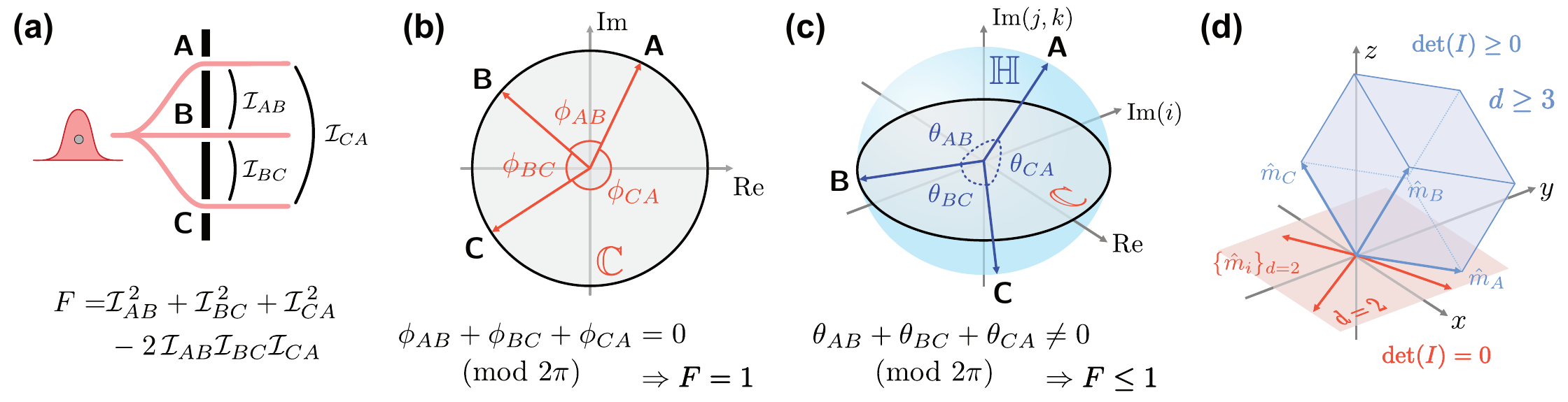}
    \caption{\textbf{Three-path interference test.} (a)~Experimental setup of the Peres test with three different paths A, B, and C. Due to Born's rule, such a setup involves three second-order interferences $\mathcal{I}_{ij}$, which can be used to construct the $F$ function. (b)~In CQM, the three phases associated to the three paths can be depicted in the unit circle with a closed phase relation leading to $F=1$. (c)~In QQM with three-dimensional phases, the closed phase relation does not hold in general leading to $F\leq 1$. (d)~Mapping the wave functions onto unit vectors $\hat{m}_i$, the $F$ function can be related to the determinant of the interference matrix $I$ yielding the square of the volume of the spanned parallelepiped. For CQM with $d=2$ (in red), all vectors lie within a plane, the spanned volume is zero, and thus $\det(I)=0$ equivalent to $F=1$. For higher $d\geq 3$, however, the $\hat{m}_i$'s span a nonzero volume with $\det (I) \geq 0$ equivalent to $F \leq1$.}
    \label{fig:R-C-H}
\end{figure*}

In a corresponding setup with three paths $A$, $B$, and $C$ as depicted in Fig.~\ref{fig:R-C-H}(a), three different pairwise interference terms $\mathcal{I}_{AB}$, $\mathcal{I}_{BC}$, and $\mathcal{I}_{CA}$ exist~\footnote{Note that there is no third-order interference $\mathcal{I}_{ABC}=0$ in single-particle correlations due to Born's rule~\cite{Sorkin:1994,Pleinert:2020}}.
In CQM, the latter are related to the respective phase differences $\phi_{AB}$, $\phi_{BC}$, and $\phi_{CA}$, which can be depicted in the unit circle of $\mathbb{C}$, see Fig.~\ref{fig:R-C-H}(b), building a cyclically ordered set with closed phase relation 
\begin{equation}\label{eq:closed-phase}
\phi_{AB}+\phi_{BC}+\phi_{CA}=0 \mod 2\pi \, .
\end{equation}
Starting from this relation and using trigonometrical transformations, Peres constructed the following function~\cite{Peres:1979}
\begin{equation}\label{eq:Peres_F}
    F = \mathcal{I}_{AB}^2 + \mathcal{I}_{BC}^2 + \mathcal{I}_{CA}^2 - 2 \, \mathcal{I}_{AB}\mathcal{I}_{BC}\mathcal{I}_{CA} \, ,
\end{equation}
which in CQM always equals one due to Eq.~\eqref{eq:closed-phase}. 
In QQM, however, the phases are determined in three-dimensional space, determined by the pure imaginary vector $\mathbf{v}$. Thus, the closed phase relation does not hold generally, see Fig.~\ref{fig:R-C-H}(c), and $F$ can become smaller than one. 
The value of $F$ can thus discriminate between CQM and QQM via~\cite{Peres:1979}
\begin{align}
\begin{split}
    F &= 1: \text{CQM is admissible,} \\
    F &< 1: \text{QQM is admissible,} 
\end{split}
\end{align}
which is called the Peres test.
By inserting $\mathcal{I}_{ij}$ from Eq.~\eqref{eq:I_ij}, $F$ can be written as a function of the probability distributions, $P_i,P_{ij}$, so that $F$ is readily accessible to experiments~\cite{Gstir:2021,Sadana:2022}.

Note that the Peres test can also be seen as a test for the non-commutativity of phases. 
Two quaternions and thus their phases commute, when their respective vector parts are parallel to each other, $\mathbf{v}_i || \mathbf{v}_j$. 
In that case, $\hat{n}_i \cdot \hat{n}_j =1$ and $\theta_i \rightarrow \phi_i$, such that the quaternionic interference terms $\mathcal{I}_{ij}^{\,\mathbb{H}}$ reduce to the complex ones $\mathcal{I}_{ij}^{\,\mathbb{C}}$ leading to $F=1$. 
Hence, commutativity goes along with CQM.

\paragraph*{Matrix formalism of the Peres test.}

To generalize the Peres test to multi-path and multi-particle interferences, we will now map the original Peres test to a matrix formalism better suited for a generalization. 

The second-order interference terms of Eq.~\eqref{eq:I_ij} can be expressed as a dot product of two real-valued unit vectors, i.e., $\mathcal{I}_{ij} = \hat{m}_i \cdot \hat{m}_j$ where $\hat{m}_i, \hat{m}_j \in \mathbb{R}^d$ with $d$ the dimension of the number system the wave functions are chosen in. 
For CQM with $\mathbb{C} \simeq \mathbb{R}^2$, for example, the map reads $\psi_i(x) = a_i + i b_i \rightarrow \hat{m}_i = (a_i,b_i)$. For a double slit with $\psi_i=\exp (i\phi_i)$, we get $\hat{m}_i = (\cos \phi_i, \sin \phi_i)$ and thus recover $\mathcal{I}_{ij} = \hat{m}_i \cdot \hat{m}_j = \cos (\phi_j - \phi_i)$.

For a setup of three paths as in the Peres test, we have three individual wave functions that are mapped onto vectors $\hat{m}_i$, which can be collected to define the matrix $M = (\hat{m}_A,  \hat{m}_B, \hat{m}_C)^T$. 
We can then construct an interference matrix containing all second-order interference terms via
\begin{equation}
I = MM^T 
=\begin{pmatrix}
1 &  \mathcal{I}_{AB}  &  \mathcal{I}_{AC}  \\
\mathcal{I}_{BA} & 1  &  \mathcal{I}_{BC}   \\
 \mathcal{I}_{CA}  &   \mathcal{I}_{CB}   &  1
\end{pmatrix} ,
\end{equation}
where $\mathcal{I}_{ii}=\hat{m}_i\hat{m}_i=1$ regardless of the number system, and $\mathcal{I}_{ij}=\mathcal{I}_{ji}$.
Now, the determinant of this interference matrix can be calculated to 
\begin{equation}\label{eq:det-I}
\det(I)  = 1- \mathcal{I}^{\,2}_{AB}- \mathcal{I}^{\,2}_{BC}-  \mathcal{I}^{\,2}_{CA} + 2 \mathcal{I}_{AB} \mathcal{I}_{BC} \mathcal{I}_{CA} = 1-F \, ,
\end{equation}
and contains Peres' $F$ function of Eq.~\eqref{eq:Peres_F}. 
This is no coincidence, since $\det(I)=0$ reveals the linear dependence of the three vectors $\hat{m}_i$. The three corresponding phases can thus be written in a closed phase relation as in Eq.~\eqref{eq:closed-phase} for CQM, which was the starting point of Peres’ construction of the $F$ function. 
Hence, we can state the test equivalently as

\begin{align}\label{eq:test-det}
\begin{split}
    \det(I) &= 0: \text{CQM is admissible,} \\
    \det(I) &> 0: \text{QQM is admissible.}
\end{split}
\end{align}
So far, we have not made any assumptions on the underlying number system and Eqs.~\eqref{eq:det-I} and~\eqref{eq:test-det} are thus valid for all dimensions. 
The dimension $d$ of the number system comes in via the dimension of the individual $\hat{m}_i$ and thus $M$ is $(3 \times d)$-dimensional. 
The dimension of $I=MM^T$, however, is always $(3 \times 3)$ containing the three second-order interference terms for all $d$.

In CQM, $M$ is thus $3\times 2\,$- dimensional and we can add to $M$ a column of zeros  such that it becomes a square matrix $\tilde{M}=(M,\mathbf{0})$. 
Note that this transformation does not change the interference matrix $I=\tilde{M}\tilde{M}^T$. 
Then, from $\det(\tilde{M})=0$ and the product rule of the determinant, one can immediately conclude $\det(I)=0$ or $F=1$ in CQM.
For higher $d$ as in QQM, however, no zero column can be added to make the above conclusion on $\det (I)$ and $F$.

\paragraph*{Multi-path Peres tests.}

Next, we use the introduced matrix formalism to introduce multi-path interference tests with an arbitrary number of paths $n$. 
Note that extensions based on closed phase relations, although feasible in principle, quickly become cumbersome due to the appearance of many trigonometric transformations. 
From now on, we will denote the number of paths $n$ in a subscript of the related functions and matrices, i.e., $F_n$, $M_n$, and $I_n$, with the previous example $F\equiv F_3$, $M\equiv M_3$, and $I\equiv I_3$.

In an $n$-path setup, there are $n$ different wave functions, each associated with a phase, which
can be mapped onto $d$-dimensional unit vectors $\{\hat{m}_i\}_{i=1,\ldots,n}$.
The generalized matrix $M_n$, built up of these vectors, becomes $(n\times d)$-dimensional, where the number of rows equals the number of paths $n$ in the setup and the number of columns equals the dimension $d$ of the investigated number system.
The general interference matrix $I_n = M_nM_n^T$ will thus be $(n\times n)$-dimensional collecting all $\binom{n}{2}$ (non-trivial) second-order interferences $\{\mathcal{I}_{ij}\}_{i,j=1,\ldots,n;i< j}$. 

As shown above for CQM in the original Peres test, a transformation $M \equiv M_3 \rightarrow \tilde{M_3}=(M_3,\mathbf{0})$ does not change the $I_3$-matrix, and we could directly conclude that $F_3=1$ for $d=2<3=n$.
Quite generally, this transformation is possible, when the number of paths $n$ is greater than the investigated dimension $d$, i.e., for $n>d$, we can write
\begin{equation*}
    M_n= \begin{pmatrix}
        \hat{m}_1 \\
        \vdots \\
        \hat{m}_n
    \end{pmatrix}_{\hspace{-1mm}n \times d} \hspace{-5mm} \rightarrow  \Tilde{M}_n = \begin{pmatrix}
        m_{1,1}   & \dots & m_{1,d}   & 0 & \dots & 0\\
        \vdots & \ddots & \vdots & \vdots & \ddots & \vdots \\
        m_{n,1} & \dots & m_{n,d} & 0 & \dots & 0 \\
    \end{pmatrix}_{\hspace{-1mm}n \times n} \, ,
    \label{eq:generalMfill}
\end{equation*}
and directly conclude $\det(I_n)= \det(\tilde{M}_n)^2=0$. 
The multi-path Peres function is thus always one, $F_n:=1-\det(I_n)=1$ for $d<n$. 
However, when $d \geq n$, there is no such transformation giving access to the calculation of the determinant $I_n$ via the determinant of $M_n$, 
such that we can not conclude $\det(I_n) = 0$. 
Underestimating the dimension results in the zeros of the matrix $\tilde{M}$ being replaced by higher-dimensional coefficients, giving the possibility of $\det(I_n) \geq 0$, and thus
$F_n \leq 1$. This indicates that the number system of dimension $d$ is not admissible, but a higher-dimensional one should be considered.

There is thus a direct connection between the number of paths $n$ used in the test and the dimensionality $d$ of the number system to be tested. 
The ranges of the generalized $F_n$ are shown below for $d=2,3,4,5$ and $n=3,4,5,6$.

\begin{center}
\begin{tabular}{c|c|c|c|c|}
\cline{2-5}
                            & $d=2$    & $d=3$  & $d=4$   & $d=5$  \\ \hline
\multicolumn{1}{|l|}{$F_3$}  & \textbf{1} & $\leq1$  & $\leq1$ & $\leq1$ \\ \hline
\multicolumn{1}{|l|}{$F_4$}  & \textbf{1} & \textbf{1} & $\leq 1$ & $\leq 1$\\ \hline
\multicolumn{1}{|l|}{$F_5$}  & \textbf{1} & \textbf{1} & \textbf{1} & $\leq 1$ \\ \hline
\multicolumn{1}{|l|}{$F_6$}  & \textbf{1} & \textbf{1} & \textbf{1} & \textbf{1} \\ \hline
\end{tabular}    
\end{center}
In particular, if we want to test between two arbitrary dimensions $d_1$ and $d_2$ with $d_1<d_2$, one has to use a multi-path test with $n\in [d_1+1, d_2]$ paths. 
For all other number of paths $n$, the related interference test is not sensitive to the relevant dimensions:
For $n < d_1+1$, both theories yield $F_n\leq 1$, while for $n > d_2$, both theories yield $F_n=1$. 
In the case of CQM ($d=2$) vs. QQM ($d=4$), future experiments can thus use $F_3 \equiv F$ but likewise $F_4$.

\paragraph{Geometric interpretation.}

In mathematical terms, the interference matrix $I_n$ constitutes a Gram matrix of the vectors $\{\hat{m}_1,\ldots,\hat{m}_n\}\in\mathbb{R}^d$; and its determinant gives the square of the volume of the $n$-parallelotope ($n$-dimensional extension of a $3D$-parallelepiped) spanned by the vectors.
This $n$-dimensional volume, built up of \emph{unit} vectors $\hat{m}_i$, ranges from $0$, if all vectors lay within an $(n-1)$-dimensional subspace, to $1$, if all vectors are linearly independent, such that $\det(I_n), F_n \in [0,1]$ for all $n\geq 3$.

In the original Peres test with $F_3$, the determinant of $I$ thus gives the square of the volume of the $3D$-parallelepiped spanned by $\hat{m}_A$, $\hat{m}_B$, $\hat{m}_C$.
For CQM, however, the $\hat{m}_i$'s are two-dimensional and lie within the $x-y$ plane, see Fig.~\ref{fig:R-C-H}(d). 
The spanned volume is thus zero and so is $\det(I)=0$ leading to $F_3\equiv F=1$ (the original Peres test).
For higher $d\geq 3$, however, the $\hat{m}_i$'s do not generally lie within a plane, see blue case in Fig.~\ref{fig:R-C-H}(d), and the spanned volume as well as $\det(I)$ might be nonzero, i.e., $F<1$.

\paragraph*{Multi-particle Peres tests.}

So far, the constructed tests are based on single-particle interference. Quantum mechanics, however, also allows for multi-particle interference in the case of indistinguishable particles~\cite{Glauber:1963,Pan:2012,Tichy:2014,Pleinert:2020}.
In the following, we introduce a multi-particle extension of the Peres test in the setup of mutually coherent particles,
where we denote the number of particles $m$ in a superscript of the functions, e.g., $F^{(m)}_n$, with the previous $F_n\equiv F^{(1)}_n$.

We specifically consider an $m$-particle wave function that is coherently spread over $n$ paths. 
The state of this wave function can be described by a tensor product of the single-particle states, and the $m$-particle Hilbert space is given by $\mathcal{H}^{m} = \bigotimes_{i=1}^{m} \mathcal{H}_i$. 
The joint probability of coincidentally detecting the state at $m$ detectors is given by the $m$th-order intensity correlation function~\cite{Glauber:1963}. 
The latter is in general determined by $n^m$ different, yet indistinguishable paths leading to in total $n^{2m}$ interference-like terms that can be sorted into interference orders up to order $2m$~\cite{Pleinert:2020}.
These interference terms are the building blocks of the Peres test.
For mutually coherent particles like an $m$-particle Fock state, we can recover all the terms by exploiting tensor products together with the matrix formalism. We obtain 
$I_{n}^{(m)} =  \bigotimes_{i=1}^m I^{(1)}_{n,i} $, where $I^{(1)}_{n,i}$ is the $n$-path single-particle interference matrix of the $i$th particle.
The multi-particle Peres test can then be defined similarly to the single-particle case via $F_{n}^{(m)} = 1 - \det(I_{n}^{(m)})$.
Here, we can make use of the identity, $\det (K \otimes L) = \det(K)^{l} \det(L)^{k}$, with $k$ and $l$ being the dimensions of the matrices $K$ and $L$ respectively. 
In our case, $k=l=n$ such that $\det(I_{n}^{(m)}) =  \prod_{i=1}^m \det (I^{(1)}_{n,i})^{n} $. 
Inserting this and using the $n$-path version of Eq.~\eqref{eq:det-I} for the $i$th particle, $F^{(1)}_{n,i} = 1 - \det (I^{(1)}_{n,i})$, we eventually obtain the multi-path and multi-particle generalization of the Peres test, i.e.,
\begin{equation} \label{eq:multiF}
F_{n}^{(m)} = 1-\prod_{i=1}^m (1- F^{(1)}_{n,i})^{n} \, .   
\end{equation}
Here, $F_{n}^{(m)}$ is directly related to the single-particle functions $F^{(1)}_n$ and the connection between the number of paths $n$ and the dimensionality $d$ of the number system can be  adopted:
\begin{align*}\label{eq:test-F-multi}
\begin{split}
    F_{n}^{(m)} &= 1: \text{a theory with $d=n-1$ is admissible,} \\
    F_{n}^{(m)} &< 1: \text{a higher-dimensional theory is admissible.}
\end{split}
\end{align*}
Note that due to this relation, the $m$-particle Peres function $F_{n}^{(m)}$ approaches faster than $F_{n}^{(1)}$ the case, where a lower-dimensional theory is admissible.

\paragraph*{Conclusion.}

In summary, we derived in a most general way the Peres test via a matrix formalism, what allows in particular for a geometrical interpretation of the test. 
We further introduced generalized Peres tests that exploit multi-path and multi-particle interference, revealing a direct relation to the dimension of the number system of quantum mechanics.
Future theoretical works have to explore, how multifaceted interference in general, e.g., using mutually incoherent particles modifies these tests and its sensitivity to higher dimensions; furthermore, future experiments have to show whether complex numbers are not only necessary but also sufficient for quantum mechanics.

\paragraph*{Acknowledgments.}
E.\.{I}.S. thanks Miguel Navascués for fruitful discussions and gratefully acknowledges financial support by the International Max Planck Research School: Physics of Light (IMPRS-PL).
J.v.Z. and M.-O.P. acknowledge funding by the Erlangen Graduate School in Advanced Optical Technologies (SAOT) by the Bavarian State Ministry for Science and Art. This work was supported by the Deutsche Forschungsgemeinschaft (DFG, German Research Foundation) – Project-ID 429529648 – TRR 306 QuCoLiMa (“Quantum Cooperativity of Light and Matter’’).

\bibliography{ref}

\end{document}